\documentclass{ws-procs975x65}

\begin{document}

\def\figsubcap#1{\par\noindent\centering\footnotesize(#1)}

\title{\uppercase{Modeling the production of intergalactic light in the
    pre-collapse phase of galaxy groups}}

\author{JOSE M.\ SOLANES and LAURA DARRIBA for THE IDILICO COLLABORATION$^*$}
\address{Departament d'Astronomia i Meteorologia and Institut de Ci\`encies del Cosmos\\
Universitat de Barcelona, C.\ Mart\'{\i} i Franqu\'es, 1, E-08028 Barcelona, Spain\\
$^*$ All members of the IDILICO collaboration are co-authors of this paper. The full list can be found at \texttt{http://www.am.ub.edu/extragalactic/idilico/people.htm}.}

\begin{abstract}
This paper reports one recent result from a set of pre-virialized
galaxy group simulations that are being used in an investigation of
measurement techniques for the quantity of intragroup light (IGL). We
present evidence that the binding energy of the stellar material
stripped from the galaxies is essentially uncorrelated with the local
mass density. This suggests that IGL detection methods based on the
distribution of luminosity perform poorly in detecting the unbound
stars.
\end{abstract}

\keywords{galaxies: interactions, intergalactic medium, methods: $N$-body simulations}

\bodymatter

\section{Introduction}
The IDILICO project is an international collaboration that studies the
assembly of galaxy aggregations from high-resolution, numerical
simulations. We have recently finished the construction of an $N$-body
model of a pre-virialized galaxy group specifically tailored for
gaining understanding of the role of gravity on structure formation
and galaxy evolution in the framework of the $\Lambda$CDM concordance
cosmology. The properties of our group model are motivated by current
theories of hierarchical galaxy formation and satisfy numerous
observational constraints, providing a realistic caricature of these
systems in which galaxies endure both continuous mass accretion from
their surroundings and frequent interactions with neighbors in a
changing environment.

Presently, our interest is focused on the galaxy mergers that occur
within collapsing groups because of the hierarchical buildup of
structure and, especially, on the accompanying production of diffuse,
low surface brightness intergalactic light (IGL). Our method is to
build and analize a large statistical sample of such mergers. Here, we
present one of the most important results we have obtained so far.

\section{Simulations}
Groups with $O(10^7)$ particles are created as uniform, spherical CDM
perturbations at $z_{\mathrm{i}}=3$. The initial overdensities are
such that the groups expand first linearly, then go through the
turnaround, and finally experience a completely non-linear collapse at
$z\sim 0$, as done in the pioneering work by
Ref.~\refcite{DGR94,GTC96}. Each group contains 20 randomly
distributed galactic halos with a Navarro-Frenk-White profile
truncated at the virial radius. A $5\%$ of the total galactic halo
mass is placed at its center in the form of cold baryons (stellar
particles). The gross structural and dynamical properties of this
component are fixed by the virialized, extended CDM envelope following
the analytical galaxy formation model by Ref.~\refcite{DS10}.

The total mass of the galactic halos is assigned according to a Monte
Carlo realization of a flat ($\alpha=-1$) Schechter mass function down
to $M/M^*=0.05$. The smallest halos, $0.05 < M/M^* < 0.1$, are assumed
to host only spheroidal stellar distributions. Above this latter mass
threshold, morphologies are randomly drawn assuming a late-type galaxy
fraction of 0.7. For early-type objects and bulges we adopt a
Hernquist profile, while late-type galaxies have exponential stellar
disks and isothermal vertical distributions. The adopted DM to stellar
particle mass ratio is 28.50. The groups' mass is set to
$M_{\mathrm{grp}}=10^{13}\,h^{-1}M_\odot$.

The evolution of structure in our simulations is traced by a procedure
that relies on an unbiased definition of the reference frame for
unbinding which is unaffected by the recursive particle removal. Our  
halo finder algorithm also allows galactic halos to experience mass
losses due to gravitational interaction with their neighbors, as well
as mass growth due to the accretion and merger of surrounding matter
and/or the re-capture of already stripped material.

\section{Discussion and Results}
The IGL consists of stars not bound to any galaxy. This physically
appealing definition is, however, not observationally feasible. One of
the handy alternatives to replace binding energy is local
density. In its simplest form, density-based definitions adopt a
limiting threshold below which stellar particles are considered
IGL. This scheme is premised on the simple idea that stars are pulled
from high to low density as they are stripped from their parent
galaxies becoming thereafter unbound. Here we make quantitative
measurements of the stellar component in our simulated mergers using
both variables to test the validity of the hypothesis that
gravitational stripping entails the production of low-density stellar
material that is unbound.
\begin{figure}[b]%
\begin{center}
 \parbox{2.3in}{\epsfig{figure=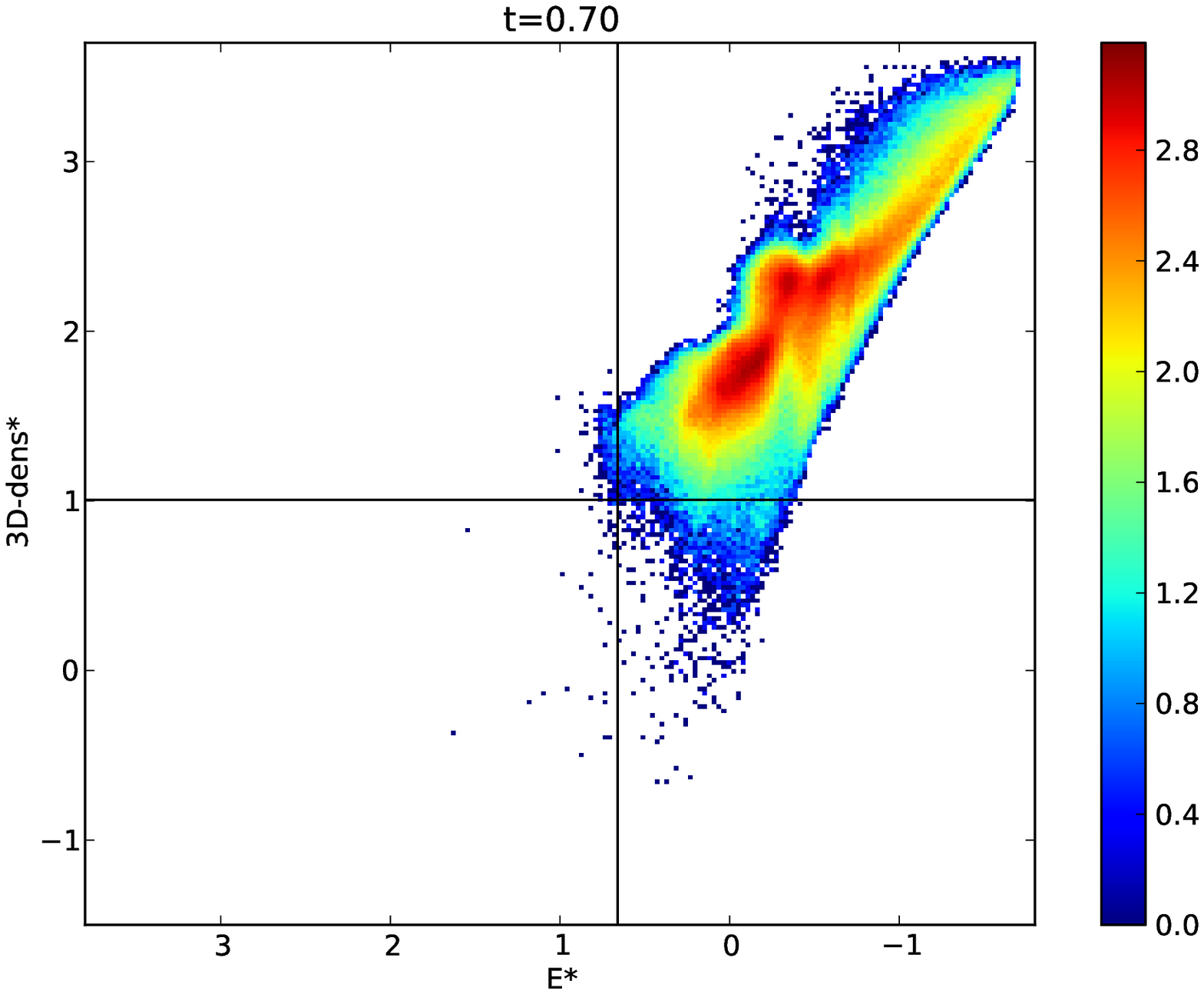,width=2.15in}}%\figsubcap{a}}
 \hspace*{4pt}
 \parbox{2.3in}{\epsfig{figure=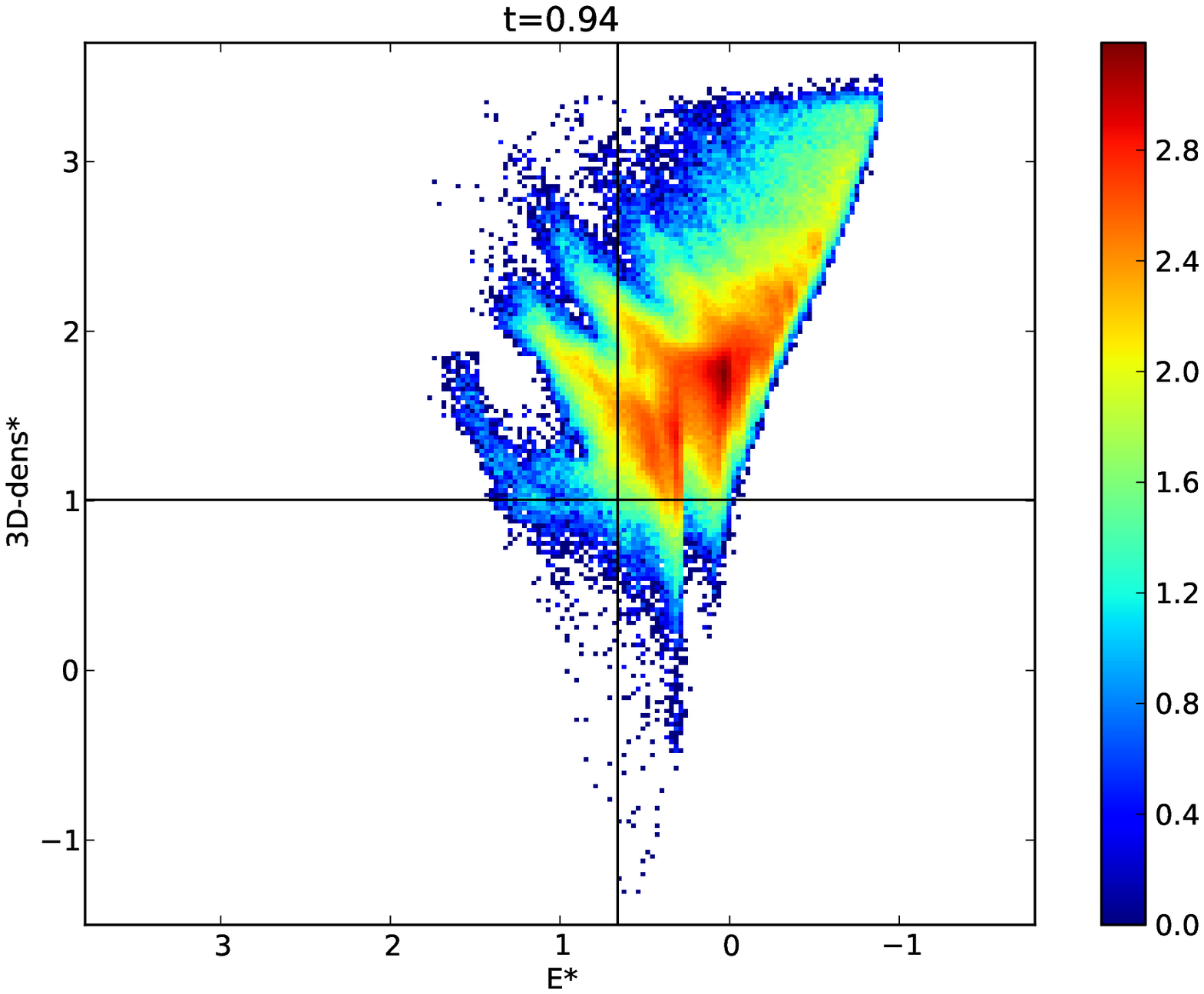,width=2.15in}}%\figsubcap{b}}
 \vspace*{10pt}
 \caption{Correlation between the (logarithm of the) 3D local density
   and binding energy of the stellar particles of a merging spiral
   galaxy. The two epochs depicted correspond to the beginning (left
   panel) and end (right) of the merger event. Note that binding
   energy grows to the right.}
\label{Fig1ab}
\end{center}
\end{figure}

Fig.~\ref{Fig1ab} shows an example of the correlation diagrams that
are inferred between the 3D local density and binding energy of the
stellar component of a given galaxy during a merger event. In this
specific case, we depict two time steps encompasing a merger between
two large disks. In the plots, we use standardized values (z-scores)
for the variables calculated from the mean and standard deviation of
the data along the entire process. Color levels show the number of
particles per pixel in logarithmic steps. In each frame, the black
vertical line indicates the z-score value corresponding to a null
binding energy (bound particles fall to the right of this line) for
the selected galaxy. The black horizontal line, on the other hand,
separates galactic (above) from IGL particles (below). The adopted
density threshold, $\rho_{\mathrm{IGL}} =
10^{-4}M_\odot\mathrm{\;pc}^{-3}$, has been determined from
observational arguments.

In the left panel of Fig.~\ref{Fig1ab} we see that before the merger
begins the binding state of the stellar particles and their local
density are reasonably related. The data delineate a broad diagonal
band in the bound-high-density quadrant (top-right), with particles in
denser regions having the tendency of being more bound and
viceversa. There is a very small fraction of luminous mass, $\sim
0.6\%$, that can be classified as diffuse according to the adopted
density threshold, but that is nevertheless bound, and a yet smaller
fraction, $\sim 0.07\%$, of unbound particles having $\rho >
\rho_{\mathrm{IGL}}$. The right panel shows the same correlation when
the merger is virtually over. According to the starting premise,
at this stage one would expect to see that a portion of the stellar
particles has become underdense and unbound, spreading over the
bottom-left quadrant of the plot. Nevertheless, this graph shows that
among the $\sim 8\%$ of the particles detached from the galaxy, very
few have local mass densities below $\rho_{\mathrm{IGL}}$, while most
of the light that would be termed diffuse (a mere $\sim 2.5\%$) is
bound anyway. Therefore, we find that in a major merger the IGL
particle candidates systematically avoid being both unbound and
underdense to the extent that, in this exemple, the fraction
of stellar light that at some point during the merger process is in the
bottom-left quadrant of the $\log\rho$ vs.\ $E$ plots never exceeds
$2\%$.

In summary, while in our pre-collapse group simulations we measure
amounts of diffuse, low-surface brightness IGL comparable to those
predicted in previous works\cite{Rud09}, we find that only a very
small fraction of it is non-galactic light. We therefore warn that
commonly used IGL detection methods based on the current position of
the group luminosity, such as surface brightness or local density
thresholds, may perform very poorly in detecting the truly unbound
stellar material.

\section*{Acknowledgements}
JMS and the IDILICO project are supported by the Subdirecci\'on
General de Proyectos de Investigaci\'on del MINECO of Spain under
grant AYA2010-18605.

\bibliographystyle{ws-procs975x65}
\bibliography{main}

\end{document}